\documentclass[prl,twocolumn,showpacs,superscriptaddress,preprintnumbers,amsmath,amssymb]{revtex4}
\usepackage{graphicx}
\usepackage{dcolumn}
\usepackage{bm}

\newcommand{\sqrts}{\ensuremath{\sqrt{s_{\textrm{\tiny NN}}}}}
\newcommand{\mub}{\ensuremath{\mu_{B}}}

\newcommand{\gev}{\ensuremath{\mathrm{\ GeV}}}

\newcommand{\mev}{\ensuremath{\mathrm{\ MeV}}}
\newcommand{\tev}{\ensuremath{\mathrm{\ TeV}}}
\newcommand{\fm}{\ensuremath{\mathrm{fm}}}

\newcommand{\changed}{}

\begin{document}

\preprint{}

\title{Hadron Formation in Relativistic Nuclear Collisions and the QCD Phase Diagram}

\author{Francesco Becattini}
\affiliation{Universit\`a di Firenze and INFN Sezione di Firenze, Firenze, Italy}

\author{Marcus Bleicher}
\affiliation{Frankfurt Institute for Advanced Studies (FIAS), Frankfurt, Germany}

\author{Thorsten Kollegger}
\affiliation{Frankfurt Institute for Advanced Studies (FIAS), Frankfurt, Germany}

\author{Tim Schuster}
\affiliation{Yale University, New Haven, CT, USA}

\author{Jan Steinheimer}
\affiliation{Nuclear Science Division, Lawrence Berkeley National Laboratory, Berkeley, California 94720, USA}

\author{Reinhard Stock}
\affiliation{Frankfurt Institute for Advanced Studies (FIAS), Frankfurt, Germany}
\affiliation{Institut f\"ur Kernphysik, Goethe-Universit\"at, Frankfurt, Germany}

\date{\today}

\begin{abstract}
{\changed We have studied particle production in ultrarelativistic nuclear collisions at CERN SPS and LHC energies and the conditions of chemical freeze-out. We have determined the effect of the inelastic reactions between hadrons occurring after hadronization and before chemical freeze-out employing the UrQMD hybrid model. The differences between the initial and the final hadronic multiplicities after the rescattering stage resemble the pattern of data deviation from the statistical equilibrium calculations. By taking these differences into account in the statistical model analysis of the data, we have been able to reconstruct the original hadrochemical equilibrium points in the $(T,\mu_{B})$ plane which significantly differ from chemical freeze-out ones and closely follow the parton-hadron phase boundary recently predicted by lattice QCD. }
\end{abstract}


\maketitle

The phases, and phase transformations of strongly interacting matter represent one of the key remaining questions of the Standard Model. It is the goal of Quantum Chromodynamics (QCD) theory to delineate a phase diagram of such matter \cite{Kapusta:1979fh}. As its most prominent feature, recent results of lattice QCD calculations \cite{Karsch2004:QGP3, Allton:2003vx, Fodor:2001pe} predict a phase transformation between confined hadrons and deconfined quarks and gluons. A parton-hadron coexistence line results, in the plane spanned by temperature $T$ and baryochemical potential $\mub$, the principal variables of a phase diagram derived from the grand canonical equilibrium thermodynamics of quarks and gluons, as considered on the lattice \cite{Karsch2004:QGP3}. The coexistence line (or phase boundary) originates, at $\mub=0 \mev$, with a temperature $T=165\pm8 \mev$, far into the nonperturbative sector of QCD. The nature of the transition is a narrow cross-over here \cite{Karsch:2011mg}. It continues, with a slight downward curvature, up to a $\mub$ of about $600 \mev$ {\changed \cite{Endrodi:2011gv, Kaczmarek:2011zz, Cea:2012ev}} perhaps featuring a critical point \cite{Schmidt:2008ev, Cheng:2008zh} whereupon the transition would become first order.

Relativistic nucleus-nucleus collisions aim to identify such features of the phase diagram \cite{Stock:2008ru}. The large collisional volume undergoes an evolution of the contained QCD matter, starting from conditions far from quark-gluon equilibrium during interpenetration of the collision partners. After a certain formation time the collisional fireball will approach quark and gluon chemical equilibrium, at least locally, and a hydrodynamic expansion evolution will set in which proceeds along a trajectory in the $(T,\mub)$ plane \cite{Kolb2004:QGP3}. {\changed With increasing collision energy these trajectories sample across this} plane toward $\mub=0 \mev$, the site of the primordial cosmological evolution, which is closely approached by recent experiments at RHIC and LHC \cite{Schutz:QM2011}. Various physics observables get formed at different stages of the evolution. They "freeze in" thus surviving the subsequent stages essentially unobliterated, and thus preserve information pertinent to various regions of the  phase diagram \cite{Stock:2008ru}. In this Letter we shall focus on results concerning the position, in $(T,\mub)$, of the parton-hadron phase boundary line. In fact we will show that hadron production data confirm the QCD results \cite{Endrodi:2011gv, Kaczmarek:2011zz} for this line.

Hadrons get formed once the expansive evolution crosses the phase boundary. Hadronization seems to be close to the chemical freeze-out point, that is the point where subsequent inelastic collisions between hadrons cease and hadronic species abundances get frozen \cite{Stock:1999hm,Heinz:1999kb}. Moreover, the resulting hadronic yield distributions, over the various species, can be understood to closely resemble grand canonical Gibbs equilibrium ensembles, from AGS to LHC energies. This observation of equilibrium \cite{Hagedorn:1965st} is exploited in the Statistical Hadronization Model (SHM) \cite{Becattini:2003wp, BraunMunzinger2004:QGP3,Andronic:2008gu} which yields a "freeze-out point" in the $(T,\mub)$ plane, for each collision geometry and energy studied. Such points are then smoothly interpolated yielding the hadro-chemical freeze-out curve \cite{Cleymans:2005xv} which is frequently shown in the QCD phase diagram.

The freeze-out curve converges towards the lattice QCD phase boundary line, at small $\mub$, thus confirming the transition temperature of about $165 \mev$ \cite{Cleymans:2005xv}. But it appears to fall well below the line toward higher values of $\mub$, a widely discussed feature \cite{McLerran:2008nn} which has, however, not been conclusively understood.

We have shown in a previous publication \cite{Becattini:2012sq} that the hadronic freeze-out curve needs revision. All previous determinations of its points in the $(T,\mub)$ plane, in the framework of SHM, have implicitly assumed that {\changed the} primordial hadro-chemical equilibrium (an intrinsic feature of the hadronization process as observed in elementary collisions \cite{becareview}) remains frozen-in throughout the final expansion phase. In other words, the chemical freeze-out point was assumed, in these analyses, to coincide with the {\em point of the latest chemical equilibrium} for the hadronic species.
This assumption turns out, on the one hand, to be realistic as far as the inelastic sector during the hadronic phase is concerned, that delivers the bulk mesonic output. On the other hand, the baryon-antibaryon annihilation and regeneration processes do not fall away with the onset of expansive cooling \cite{Becattini:2012sq, Steinheimer:2012rd, Rapp:2000gy, Greiner:2000tu}. Their final effect consists of a considerable distortion of the initial, post-hadronization equilibrium yield distribution, in the antibaryon and baryon sector \cite{Bass:2000ib}. We have shown \cite{Becattini:2012sq} that such distortions {\changed affect the outcome of the SHM analysis,} universally {\changed leading} to a split between the latest hadrochemical equilibrium point (LHCE) and the chemical freeze-out point with a downward shift of the latter \cite{Cleymans:2005xv}, and to unsatisfactory SHM fits. In fact these effects have recently also been noticed with the first LHC hadron production data of ALICE \cite{ALICE:2012iu, ALICE:2012QM}, which resulted in recognition of a "non-thermal proton to pion 
production ratio" \cite{Antinori:2011us}.

{\changed We shall demonstrate in this Letter that an appropriate correction of the SHM predictions for the hadronic multiplicities leads to a revised LHCE curve, over the $\mub$ domain covered by the data gathered at the SPS \cite{Anticic:2010mp} and at current LHC \cite{ALICE:2012QM}  energies. } We employ {\changed modification } factors derived from analysis of the cascade phase effects employing the recent hybrid version of the microscopic transport model UrQMD \cite{Bleicher:1999xi, Petersen:2008dd}, that agree, at the SPS energy, with results of a former study by Rapp and Shuryak \cite{Rapp:2000gy} that employed a blast wave expansion model including detailed balance of baryon-antibaryon interaction. The resulting LHCE points will be shown to coincide with the recent predictions \cite{Endrodi:2011gv, Kaczmarek:2011zz} of the parton-hadron boundary line from lattice QCD at finite baryochemical potential.

We illustrate in Fig.1 the effects of the final hadron/resonance expansion phase on the observed yield distributions, as derived from the UrQMD hybrid version \cite{Petersen:2008dd}. It features a 3+1 dimensional hydrodynamic expansion during the high density stage, terminated by the Cooper-Frye hadron formation mechanism. In order to account for the considerable time dilation that occurs toward large rapidity we have changed the "isochronous" procedure of \cite{Petersen:2008dd}. We hadronize \cite{Li:2008qm} in successive transverse slices, of thickness $\Delta(z)=0.2\ \fm$, whenever all fluid cells of that slice fall below a "critical" energy density, assumed here to be $0.8 \gev/ \fm^{3}$. We thus achieve a rapidity independent freeze-out temperature. The hadron distribution can be examined at this stage, emitting into vacuum. Alternatively, the UrQMD cascade expansion stage is attached to the Cooper-Frye output, as an "afterburner". The effect of this stage can be quantified by modification factors, for each hadron multiplicity, $M = N(Hydro+Aft)/N(Hydro)$. These factors are shown in Fig.1(top panel). It illustrates the results obtained for central Pb+Pb collisions at SPS energies, $\sqrts=17.3 \gev, 8.7 \gev$ 
and $7.6 \gev$ \cite{Becattini:2012sq}, and for the present top LHC energy, $\sqrts=2.76\tev$ \cite{Steinheimer:2012rd}. At the SPS energies we see the bulk hadron output relatively unaffected by the afterburner, including the $\Xi$, $\Omega$ and $\bar \Omega$ yields. Whereas the other antibaryons, $\bar p$, $\bar \Lambda$ and $\bar \Xi$ are showing significant suppressions ranging from 50 to 25\%. At LHC energy the modification factors of baryons and antibaryons in Fig.1 become approximately equal as is to be expected in view of the particle-antiparticle symmetry prevailing here (with $\mub$ close to zero). The suppression pattern differs, in detail, from the pattern at SPS energy. It appears to be restricted to the $p$, $\bar p$, $\Xi$ and $\bar \Xi$ yields whereas the $\Lambda$, $\Omega$ and their antiparticles exhibit influences of a possible dynamical regeneration (see ref \cite{Steinheimer:2012rd} for discussion).

\begin{figure}[tbp]
 \includegraphics[width=8cm]{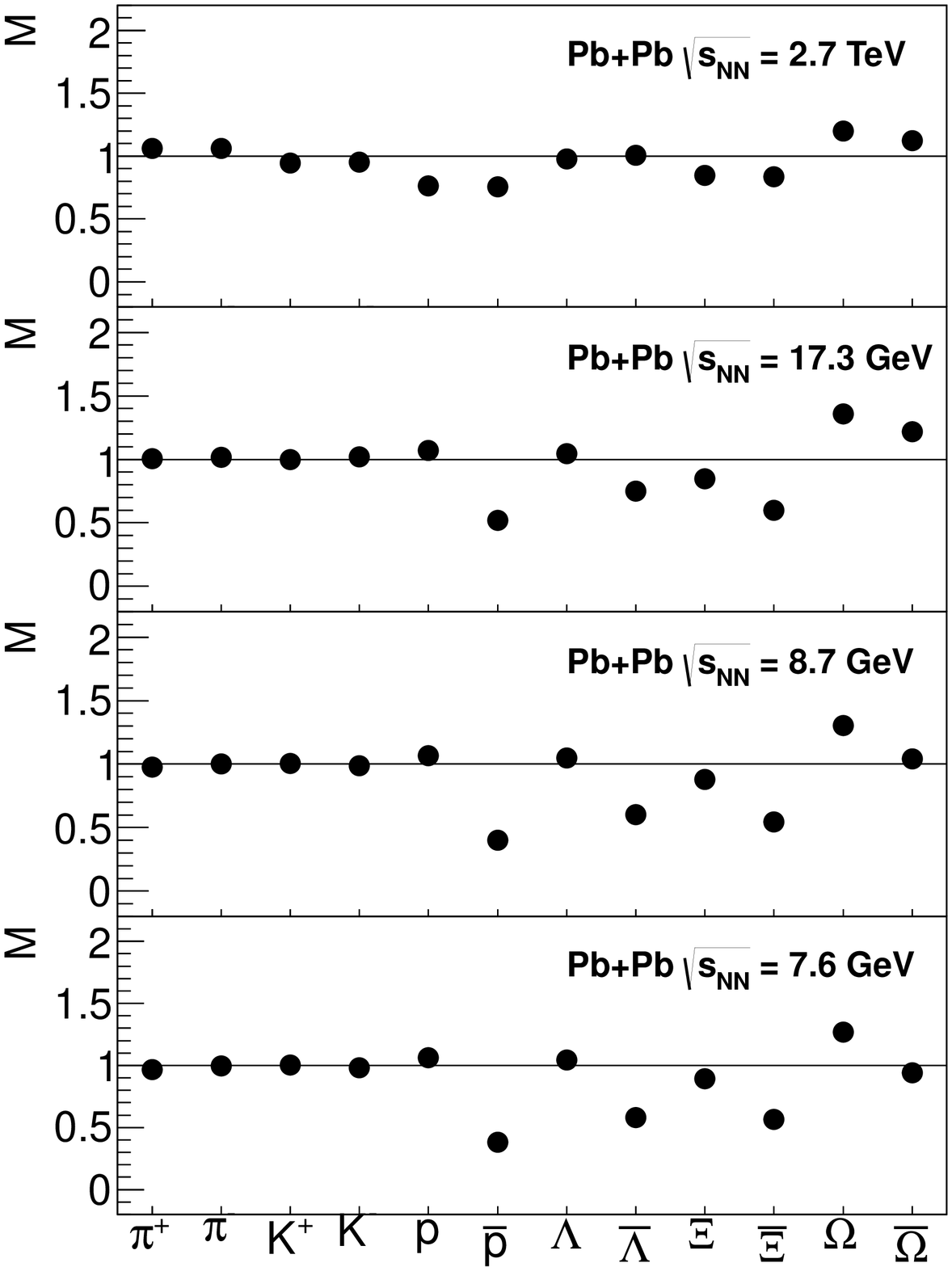}
 \includegraphics[width=8cm]{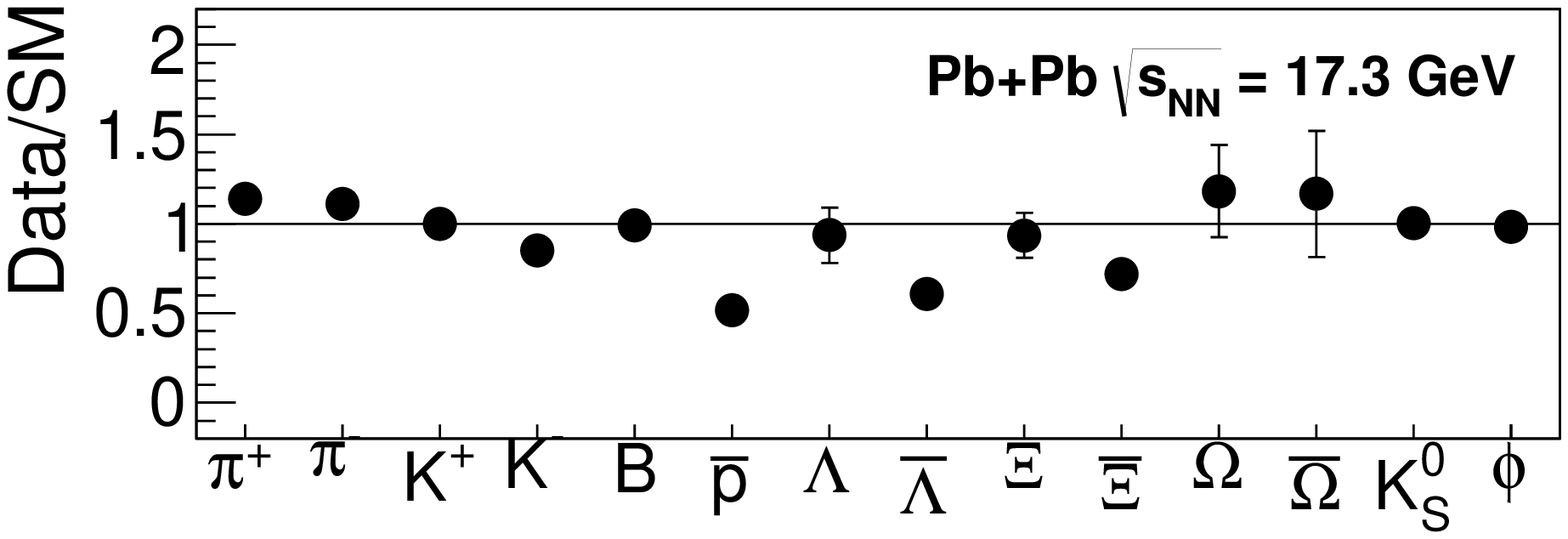}
 \label{fig:ModificationFactors}
 \caption{Top panel: Modification factors from UrQMD between particle multiplicities at hadronization and after the hadronic cascade afterburner stage for Pb-Pb collisions at $\sqrts = 2.7 \ \tev$, $17.3 \ \gev$, $8.7\ \gev$ and $7.6\ \gev$. Bottom panel: Ratio between central Pb-Pb collisions data at $\sqrts = 17.3 \gev$ and a statistical model fit excluding antibaryons \cite{Becattini:2012sq}, for which the effect of the hadronic stage is largest.}
\end{figure}

Within the above model considerations the annihilation and/or regeneration effects inflict distortions of the initial equilibrium yield distributions
imprinted into the subsequent cascade phase by the grand canonical Cooper-Frye formalism. It is important to demonstrate that a quantitatively similar distortion pattern governs the experimentally observed hadron multiplicity data. To this end we have performed a SHM analysis of the NA49 hadron yield data \cite{Anticic:2010mp} for central Pb+Pb collisions at $17.3\gev$ excluding the most affected antibaryon species $\bar p$, $\bar \Lambda$ and $\bar \Xi$ from the fit procedure (see ref \cite{Becattini:2012sq} for detail). The result is shown in the bottom panel of Fig.1. It shows the ratios of data relative to {\changed SHM} predictions, for all species. We note that the bulk hadron data are well reproduced whereas the yields of $\bar p$, $\bar \Lambda$ and $\bar \Xi$ exhibit a strong suppression relative to the {\changed SHM} equilibrium distribution of multiplicities. The pattern quite closely resembles the UrQMD prediction in the upper panel of Fig.1. 

These observations lead to the idea to employ the UrQMD "survival factors" on face value: as {\changed modification } factors employed in the SHM analysis that aims at constructing a LHCE curve. Such an analysis is shown in Fig.2, applied to recent LHC ALICE data \cite{ALICE:2012QM} for the 20\% most central Pb+Pb collisions at $2.76\tev$. The top panel gives the standard SHM fit which is unsatisfactory, the bulk pion and kaon yields being missed at the cost of accounting for the baryon sector. Similar results have been obtained in ref.~\cite{Andronic:2012dm}. The bottom panel shows the analysis with {\changed modification } factors (the survival factors from UrQMD in Fig.1) applied to the SHM fit procedure. It yields a $(T,\mub)$ point at $(166 \mev,2 \mev)$, with improved $\chi^{2}$.

We repeated this analysis with the NA49 data for 5\% centrality selected Pb+Pb collisions \cite{Anticic:2010mp} at $\sqrts = 7.6, 8.7$ and $17.3 \gev$. All obtained {\changed SHM} parameters are gathered in Tab.1. We do not include the RHIC data in the present analysis because the hadron multiplicities are not yet systematically corrected for feed-down into p, $\bar p$, $\Lambda$ and $\bar \Lambda$ from weak decays (an effect that counteracts the annihilation losses), and because a previous analysis \cite{Manninen:2008mg} has met with considerable difficulties in cross-normalizing between the 3 experiments.  More data are forthcoming from the RHIC beam energy scan program \cite{Mohanty:2011nm} which can be used to systematically extend the present SHM analysis in the future.

\begin{figure}[htbp]
 \includegraphics[width=8.3cm]{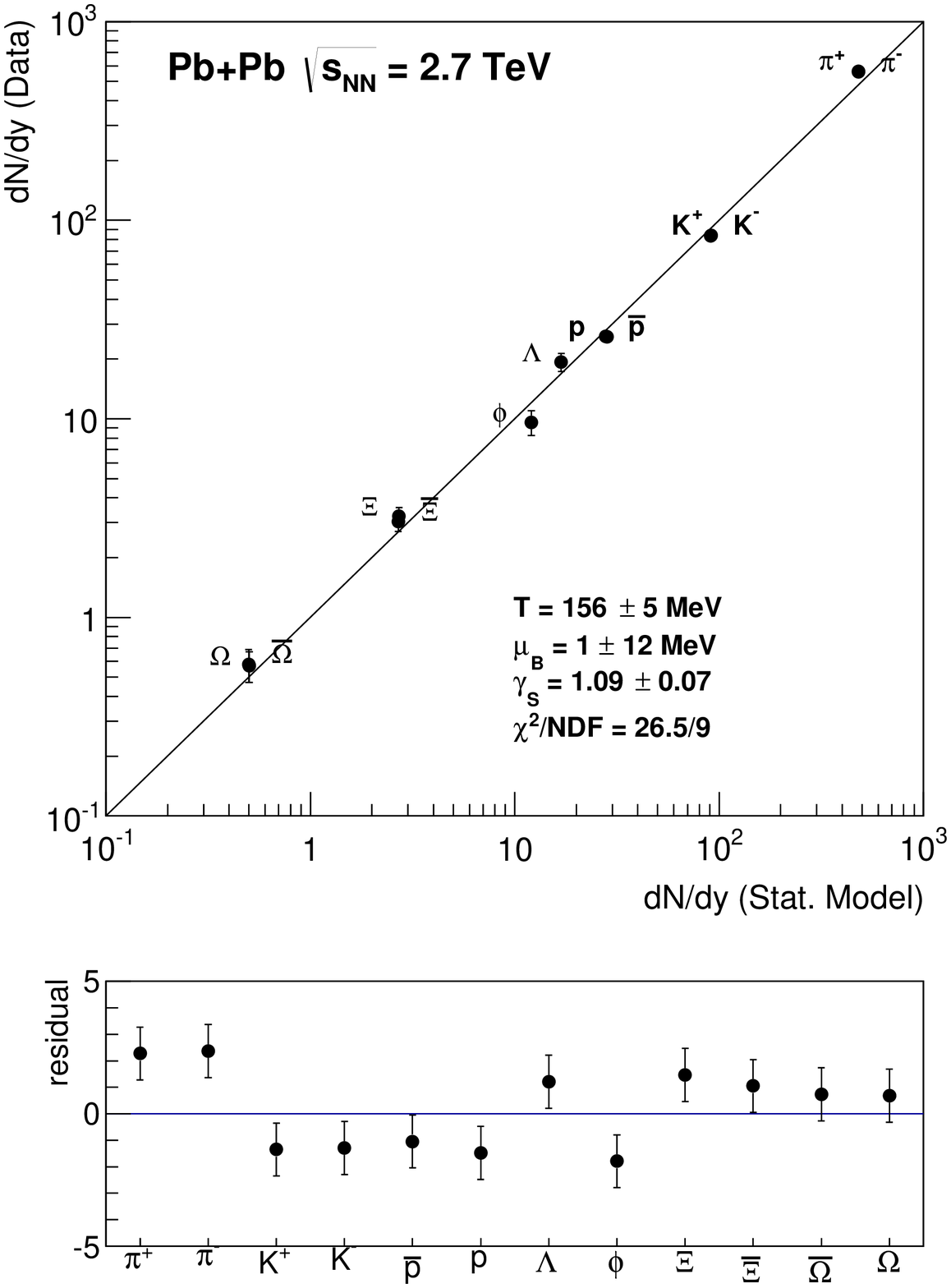}
 \includegraphics[width=8.3cm]{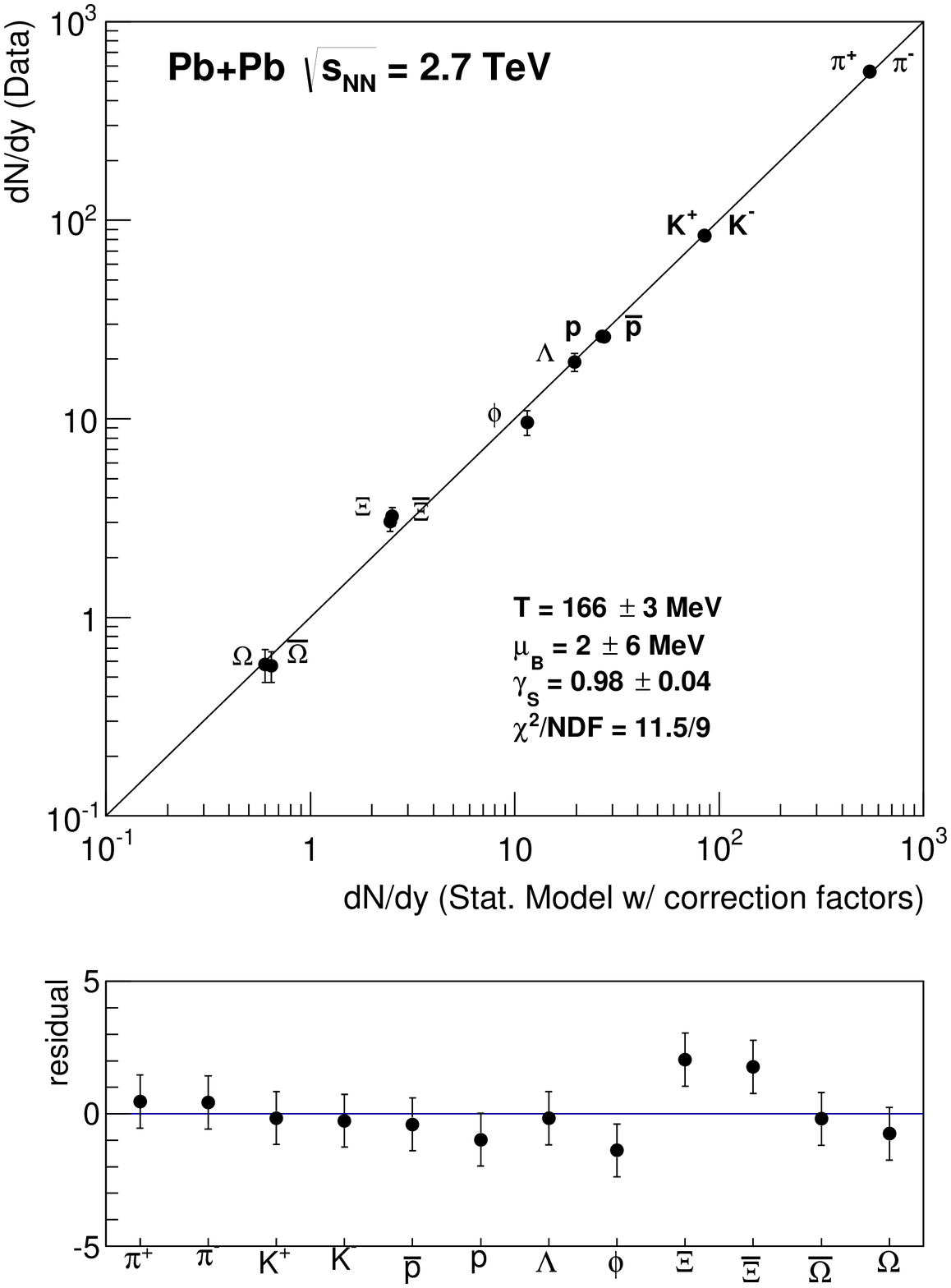}
 \label{fig:LHCFits}
 \caption{Statistical model fits to preliminary ALICE data \cite{ALICE:2012QM} for $20\%$ central Pb-Pb collisions at $\sqrts = 2.7 \ \tev$ (a) and to the same data but with modification factors from UrQMD applied in the statistical model fits (b).}
\end{figure}

Fig. 3 shows our principal result. The four obtained LHCE points are inserted into a phase diagram obtained recently by QCD lattice calculations at finite baryochemical potential, by Endrodi et al.\cite{Endrodi:2011gv}. Similar results can be found in \cite{Kaczmarek:2011zz}. The authors distinguish two different determinations of the position of the critical curve, based on the chiral condensate $\langle \bar \Psi \Psi \rangle$  and on the strange quark susceptibility $\chi_{S}/T^{2}$, respectively. The resulting two close curves $T_{\mathrm{C}} \left( \mub \right)$ cover baryochemical potentials up to $600\mev$. The LHCE points follow the latter theoretical choice for the parton-hadron transition line. They are listed in Tab.1. One generally observes an upward shift of $T$ from chemical freeze-out {\changed (as analysed in the "conventional" SHM approach \cite{Becattini:2003wp, BraunMunzinger2004:QGP3, Andronic:2008gu, Cleymans:2005xv}) to the LHCE points obtained} upon application of the UrQMD {\changed modification} factors. Except at $8.7 \gev$ it is accompanied by an improved $\chi^{2}$.

 \begin{figure}[htbp]
 \includegraphics[width=\columnwidth]{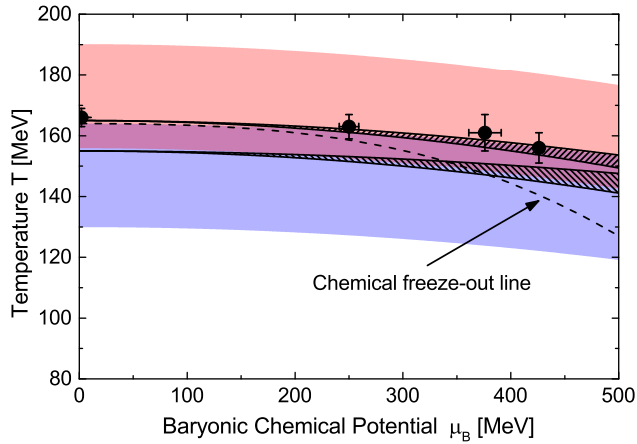}
 \label{fig:PhaseDiagram}
  
\caption{ {\changed Phase diagram of strongly interacting matter in the $\left( T, \mu_{B} \right)$ plane with predictions from lattice QCD calculations: the upper solid line is the critical temperature defined  through strange quark susceptibility, the lower one defined through the chiral condensate)  \cite{Endrodi:2011gv, Kaczmarek:2011zz} and the coloured areas represent the widths of the crossover transitions. The dashed line is the chemical freeze-out line \cite{Andronic:2008gu, Cleymans:2005xv} whilst the reconstructed original chemical equilibrium points in this work are shown as closed circles.} }
  
\end{figure}
 
 A brief reflection is in order here concerning our employ of the UrQMD hybrid transport model \cite{Petersen:2008dd}. Its account for the final hadron/resonance cascade evolution does not include the reverse of the annihilation processes, such as $p + \bar p$ to 5 pions, that could modify the survival factors \cite{Rapp:2000gy, Pan:2012ne}. This is a general feature of all existing microscopic transport models. To account for such reverse reaction channels and establish, in principle, effects of detailed balance, one has to employ analytic models of fireball expansion \cite{Rapp:2000gy, Pan:2012ne}. These, in turn, imply thermodynamically homogeneous collision volumes that miss the effects of local and surface density fluctuations. They thus overestimate both the annihilation and, in particular, the regeneration rates which scale with the fifth power of the density. Most remarkably, the main prediction \cite{Rapp:2000gy, Pan:2012ne} is, again, a net loss of about 50\% in the $\bar p$ yield, similar to {\changed the UrQMD results shown in} Fig.1.
 
In summary we have demonstrated that the semi-empirical freeze-out curve from SHM analysis of the hadronic multiplicity data at various incident energies does not coincide with the latest point at which hadrons are in chemical equilibrium following the parton-hadron conversion. To obtain the latter, one has to correct for abundance changes originating in the course of the hadron/resonance expansion that ends the dynamical evolution of A+A collisions. We have quantified those changes by survival factors obtained from the UrQMD transport model. Significant modifications are restricted to the baryon-antibaryon sector, and their detailed pattern depends on the incident energy. We have chosen to take account of the cascade phase effects {\changed by employing the survival factors to correct the SHM predictions. At the four energies considered here the resulting last equilibrium points coincide with} the lattice QCD phase boundary line, thus resolving a long standing problem. {\changed Our results give the first confirmation of the recent lattice QCD predictions \cite{Endrodi:2011gv, Kaczmarek:2011zz} at finite baryochemical potential, thus establishing the parton-hadron-coexistence line in the QCD phase diagram. } It will be very interesting to continue this analysis further upward in $\mub$ (and thus downward in incident energy).

\begin{table}[hbtp]
 \begin{tabular}{c|c|c|c|c}
 & $T (\mev)$ & $\mub (\mev)$ & $\gamma_{S}$ & $\chi^{2}/NDF$ \\
 \hline
 \hline
 \multicolumn{5}{l}{Pb-Pb 20\% central $\sqrts = 2.7 \ \tev$} \\
 \hline
 Std. fit  & $ 156 \pm 5$ & $ 1 \pm 12$ & $ 1.09 \pm 0.07$ & $26.5/9$ \\
 Mod. fit  & $ 166 \pm 3$ & $ 2 \pm 6$ & $ 0.98 \pm 0.04 $ & $11.5/9$ \\
 \hline
 \hline
 \multicolumn{5}{l}{Pb-Pb 5\% central $\sqrts = 17.3 \ \gev$} \\
 \hline
 Std. fit  & $ 151 \pm 4$ & $ 266 \pm 9$ & $ 0.91 \pm 0.05$ & $26.9/11$ \\
 Mod. fit  & $ 163 \pm 4$ & $ 250 \pm 9$ & $ 0.83 \pm 0.04$ & $20.4/11$ \\
 \hline
 \hline
 \multicolumn{5}{l}{Pb-Pb 5\% central $\sqrts = 8.7 \ \gev$} \\
 \hline
 Std. fit  & $ 148 \pm 4$ & $ 385 \pm 11$ & $ 0.78 \pm 0.06$ & $17.9/9$ \\
 Mod. fit  & $ 161 \pm 6$ & $ 376 \pm 15$ & $ 0.72 \pm 0.06$ & $25.9/9$ \\
 \hline
 \hline
 \multicolumn{5}{l}{Pb-Pb 5\% central $\sqrts = 7.6 \ \gev$} \\
 \hline
 Std. fit  & $ 140 \pm 1$ & $ 437 \pm 5$ & $ 0.91 \pm 0.01$ & $22.4/7$ \\
 Mod. fit  & $ 156 \pm 5$ & $ 426 \pm 4$ & $ 0.81 \pm 0.00$ & $14.7/7$ \\
 \hline
 \hline
 \end{tabular}
 \label{tab:FitResults}
 \caption{Results of the statistical model fits to LHC and SPS data.}
\end{table}

Acknowledgements.
We express our gratitude to the ALICE and NA49 Collaborations for providing their preliminary data. This work was supported by the Hessian LOEWE initiative through HIC for FAIR, by the Istituto Nazionale di Fisica Nucleare (INFN) and in part by the U.S. DOE under grant number DE‐SC004168. We are also grateful to the Center for Scientific Computing (CSC) at Frankfurt and to the INFN Sezione di Firenze for providing the computing resources. JS and TS were supported in part by the Alexander von Humboldt Foundation as Feodor Lynen Fellows.


\begin{thebibliography}{99}

\bibitem{Kapusta:1979fh} 
  J.~I.~Kapusta,
  Nucl.\ Phys.\ B {\bf 148}, 461 (1979).
  
\bibitem{Karsch2004:QGP3}
  F.~Karsch and E.~Laerman,
  {\it Quark-Gluon Plasma 3} (World Scientific 2004) p.1.
 
  
\bibitem{Allton:2003vx} 
  C.~R.~Allton {\it et al.},
  Phys.\ Rev.\ D {\bf 68}, 014507 (2003).

\bibitem{Fodor:2001pe} 
  Z.~Fodor and S.~D.~Katz,
  JHEP {\bf 0203}, 014 (2002).

\bibitem{Karsch:2011mg} 
  F.~Karsch,
  J.\ Phys.\ G {\bf 38}, 124098 (2011); 
  S.~Borsanyi {et al.},
  {\it ibidem} 124101.

\bibitem{Endrodi:2011gv} 
  G.~Endrodi, Z.~Fodor, S.~D.~Katz and K.~K.~Szabo,
  JHEP {\bf 1104}, 001 (2011).
  
\bibitem{Kaczmarek:2011zz} 
  O.~Kaczmarek {\it et al.},
  Phys.\ Rev.\ D {\bf 83}, 014504 (2011).


\bibitem{Cea:2012ev} 
  P.~Cea, L.~Cosmai, M.~D'Elia, A.~Papa and F.~Sanfilippo,
  Phys.\ Rev.\ D {\bf 85}, 094512 (2012);
  P.~Cea, L.~Cosmai, M.~D'Elia, A.~Papa and F.~Sanfilippo,
  PoS LATTICE {\bf 2012}, 067 (2012).



  
\bibitem{Schmidt:2008ev} 
  C.~Schmidt [for RBC-Bielefeld Collaboration],
  J.\ Phys.\ G {\bf 35}, 104093 (2008).

\bibitem{Cheng:2008zh} 
  M.~Cheng {\it et al.},
  Phys.\ Rev.\ D {\bf 79}, 074505 (2009).
  
\bibitem{Stock:2008ru} 
  R.~Stock,
  {\it Landolt-Boernstein I 21A} (Springer 2009) ch. 7
  [arXiv:0807.1610 [nucl-ex]].

\bibitem{Kolb2004:QGP3}
  P.~Kolb and U.~Heinz,
    {\it Quark-Gluon Plasma 3} (World Scientific 2004) p. 634.
  
\bibitem{Schutz:QM2011}
  see Proceedings of Quark Matter 2011, editors Y.~Schutz and U.~A.~Wiedemann,
  J.~Phys.~G38, 120301 (2011).

\bibitem{Stock:1999hm} 
  R.~Stock,
  Phys.\ Lett.\ B {\bf 456}, 277 (1999); 

\bibitem{Heinz:1999kb} 
  U.~W.~Heinz,
  Nucl.\ Phys.\ A {\bf 661}, 140 (1999)
  [nucl-th/9907060].

  J.~R.~Ellis and K.~Geiger,
  Phys.\ Rev.\ D {\bf 54}, 1967 (1996).
   
\bibitem{Hagedorn:1965st} 
  R.~Hagedorn,
  Nuovo Cim.\ Suppl.\  {\bf 3}, 147 (1965).

\bibitem{Becattini:2003wp} 
  F.~Becattini {\it et al.},
  Phys.\ Rev.\ C {\bf 69}, 024905 (2004).

\bibitem{BraunMunzinger2004:QGP3}
  P.~Braun-Munzinger, K.~Redlich and J.~Stachel,
  {\it Quark-Gluon Plasma 3} (World Scientific 2004) p. 491.

\bibitem{Andronic:2008gu} 
  A.~Andronic, P.~Braun-Munzinger and J.~Stachel,
  Phys.\ Lett.\ B {\bf 673}, 142 (2009).
  
\bibitem{Cleymans:2005xv} 
  J.~Cleymans, H.~Oeschler, K.~Redlich and S.~Wheaton,
  Phys.\ Rev.\ C {\bf 73}, 034905 (2006).

\bibitem{McLerran:2008nn} 
  L.~McLerran,
  Int.\ J.\ Mod.\ Phys.\ A {\bf 25}, 5847 (2010);
  A.~Andronic {\it et al.},
  Nucl.\ Phys.\ A {\bf 837}, 65 (2010).

\bibitem{Becattini:2012sq} 
   F.~Becattini {\it et al.},
   Phys.\ Rev.\ C {\bf 85}, 044921 (2012).
   
\bibitem{becareview}   
  F.~Becattini and R.~Fries in {\em Relativistic heavy ion physics}, Landolt-B\"ornstein 1-23
  Springer-Verlag and arXiv:0907.1031;  F.~Becattini, arXiv:0901.3643 [hep-ph].

\bibitem{Steinheimer:2012rd} 
  J.~Steinheimer, J.~Aichelin and M.~Bleicher,
  Phys.\ Rev.\ Lett.\  {\bf 110}, 042501 (2013).

\bibitem{Rapp:2000gy} 
  R.~Rapp and E.~V.~Shuryak,
  Phys.\ Rev.\ Lett.\  {\bf 86}, 2980 (2001).

\bibitem{Greiner:2000tu} 
  C.~Greiner and S.~Leupold,
  J.\ Phys.\ G {\bf 27}, L95 (2001).
  
\bibitem{Bass:2000ib} 
  S.~A.~Bass and A.~Dumitru,
  Phys.\ Rev.\ C {\bf 61}, 064909 (2000).

\bibitem{ALICE:2012iu}
  B.~Abelev {\it et al.}  [ALICE Collaboration],
  arXiv:1208.1974 [hep-ex].
  
\bibitem{ALICE:2012QM}
  L.~Milano~for~the~ALICE~collaboration, Preliminary Data, Quark Matter 2012 Conference.
  
\bibitem{Antinori:2011us} 
  F.~Antinori,
  J.\ Phys.\ G {\bf 38}, 124038 (2011).

\bibitem{Anticic:2010mp} 
  T.~Anticic {\it et al.}  [NA49 Collaboration],
  Phys.\ Rev.\ C {\bf 83}, 014901 (2011);
  NA49 Data Compilation https://edms.cern.ch/document/1075059/1; $\bar{p}$ data: NA49 preliminary.


\bibitem{Bleicher:1999xi} 
  M.~Bleicher {\it et al.},
  J.\ Phys.\ G {\bf 25}, 1859 (1999).
  
\bibitem{Petersen:2008dd} 
  H.~Petersen {\it et al.},
  Phys.\ Rev.\ C {\bf 78}, 044901 (2008).


\bibitem{Li:2008qm} 
  Q.~-F.~Li {\it et al.},
  Phys.\ Lett.\ B {\bf 674}, 111 (2009).

  

\bibitem{Andronic:2012dm} 
  A.~Andronic, P.~Braun-Munzinger, K.~Redlich and J.~Stachel,
  arXiv:1210.7724 [nucl-th].

\bibitem{Manninen:2008mg} 
  J.~Manninen and F.~Becattini,
  Phys.\ Rev.\ C {\bf 78}, 054901 (2008).
  
\bibitem{Mohanty:2011nm} 
  B.~Mohanty [STAR Collaboration],
  J.\ Phys.\ G {\bf 38}, 124023 (2011).


\bibitem{Pan:2012ne} 
  Y.~Pan and S.~Pratt,
  arXiv:1210.1577 [nucl-th].
  
\end{thebibliography}
\end{document}